\documentclass[prl,showpacs,twocolumn,floatfix,superscriptaddress]{revtex4}
\usepackage{times,amsmath,amsfonts,amssymb,latexsym}

\usepackage{graphicx,epsf}
\usepackage{bm}

\newcommand{\ket}[1]{| #1 \rangle}

\newcommand{\be}{\begin{equation}}

\newcommand{\ee}{\end{equation}}

\newcommand{\ba}{\begin{eqnarray}}

\newcommand{\ea}{\end{eqnarray}}

\newcommand{\ignore}[1]{}

\def\CC{{\rm\kern.24em \vrule width.04em height1.46ex depth-.07ex

    \kern-.30em C}}

\def\P{{\rm I\kern-.25em P}}

\def\RR{{\rm

         \vrule width.04em height1.58ex depth-.0ex

         \kern-.04em R}}

\def\bbbc{{\mathchoice {\setbox0=\hbox{$\displaystyle\rm C$}\hbox{\hbox

to0pt{\kern0.4\wd0\vrule height0.9\ht0\hss}\box0}}

{\setbox0=\hbox{$\textstyle\rm C$}\hbox{\hbox

to0pt{\kern0.4\wd0\vrule height0.9\ht0\hss}\box0}}

{\setbox0=\hbox{$\scriptstyle\rm C$}\hbox{\hbox

to0pt{\kern0.4\wd0\vrule height0.9\ht0\hss}\box0}}

{\setbox0=\hbox{$\scriptscriptstyle\rm C$}\hbox{\hbox

to0pt{\kern0.4\wd0\vrule height0.9\ht0\hss}\box0}}}}

\def\bbbz{{\mathchoice {\hbox{$\sf\textstyle Z\kern-0.4em Z$}}

{\hbox{$\sf\textstyle Z\kern-0.4em Z$}}

{\hbox{$\sf\scriptstyle Z\kern-0.3em Z$}}

{\hbox{$\sf\scriptscriptstyle Z\kern-0.2em Z$}}}}

\begin{document}
\title{Fidelity analysis of topological quantum phase transitions}
\author{Damian F. Abasto} \email{abasto@usc.edu}
\affiliation{Department of Physics and Astronomy, Center for Quantum Information Science\&Technology, University of Southern California, Los Angeles, CA 90089-0484}
\author{Alioscia Hamma} \email{ahamma@perimeterinstitute.ca}
\affiliation{Perimeter Institute for Theoretical Physics, 31 Caroline Street N, Waterloo, ON N2L 2Y5 Canada}
\author{Paolo Zanardi$^{1,}$}
\affiliation{
Institute for Scientific Interchange, Viale
Settimio Severo 65, I-10133 Torino, Italy
}

\date{April 25, 2008}

\begin{abstract}
We apply the fidelity metric approach to analyze two recently introduced models that exhibit a quantum phase transition to a topologically ordered phase. 
These quantum models have a known connection to classical statistical mechanical models; we exploit this mapping to obtain  the scaling of the fidelity metric tensor near criticality. 
The topological phase transitions manifest themselves in divergences of the fidelity metric across the phase boundaries. These results provide evidence that the fidelity approach
is a valuable tool to investigate novel phases lacking a clear characterization
 in terms of local order parameters.

\end{abstract}


\maketitle
{\em Introduction.---} This is an exciting period for condensed matter physics, when novel phases of matter that defy traditional understanding are being observed and predicted. Examples include topological phases \cite{Wen} which cannot be described by Landau-Ginzburg-Wilson paradigm \cite{gold}. Absence of local order parameters and symmetry breaking mechanisms are among the most remarkable features of these systems. These novel phases arise, for example, in collective phenomena exhibited in strongly correlated systems of two dimensional electrons at very low temperature, like in the fractional quantum Hall effect \cite{Tsui,Laughlin}. In such systems, the motion of electrons is highly constrained, and the fluctuations are entirely quantum in nature. In this situation Landau's theory, which is essentially a theory of classical order, can fail. \\
\indent It is compelling to find new ways to analyze such phases. Using tools from quantum information, it has been possible to characterize topological order using the concept of topological entropy \cite{enttopo}-\cite{enttopo3}. Here, we call for a new information-theoretic tool for studying quantum phase transitions (QPTs) \cite{QPT} to topological phases. The new notion is the fidelity of ground states, whose role in the study of QPTs has been developed in \cite{F 0}-\cite{F 23}. The basic idea is that near a quantum critical point there is a drastic enhancement in the degree of distinguishability between two ground states, corresponding to slightly different values of the parameter space that defines the hamiltonian. This distinguishability can be quantified by the fidelity, which for pure states reduces to the amplitude of inner product or overlap. This approach is suitable for detecting QPTs and analyzing topological phases, since the method does not rely on constructing an order parameter, nor on the symmetries of the system. The overlap of two nearby ground states is a global quantity of the system that does not depend on local features like the existence of a local order parameter. Therefore, it should contain all the information that describes topological order. The capability of fidelity to spot a topological QPT has been shown in \cite{num} by numerical analysis. Since topological order is a property of the ground state wave-function alone, knowledge of the ground state of the system is sufficient in order to carry out this analysis.\\
\indent In this work we analyze two models that exhibit topological order: hamiltonians with a stochastic matrix form \cite{SMD} and the quantum eight-vertex model \cite{8V1}. Both present a transition from a non-topologically ordered phase to a topological phase. Moreover, they exhibit a close connection to classical statistical models. Indeed, the fidelity and its second derivative, the so-called fidelity metric, are related to the partition function \cite{TNrep} and correlation functions of the corresponding classical model, respectively. \\
{\em Topological QPT in stochastic matrix form hamiltonians.---} In this section we apply the fidelity approach to analyze the quantum phase transition to a topologically ordered phase for a particular case of hamiltonians that exhibit a stochastic matrix form decomposition \cite{SMD}. In \cite{topo1} the authors showed their model had a transition from a magnetically ordered state to a topological ordered phase, with the topological entropy having a jump to a nonzero value at the transition. We now apply the fidelity approach to this topological QPT and find the scaling of the fidelity metric near criticality. Let us start by briefly reviewing the model. Given a square lattice with periodic boundary conditions and spins  $1/2$ on the bonds, consider the following hamiltonian:
\begin{eqnarray}
H &=& -\lambda_{0}\sum_p B_p-\lambda_1\sum_s\ A_s+\lambda_1\sum_s e^{-\beta\sum_{i\in s}\hat{\sigma}_i^z} \nonumber\\
&=& H_{\mathrm{Kitaev}}+\lambda_1\sum_s e^{-\beta\sum_{i\in s}\hat{\sigma}_i^z}\label{H1},
\end{eqnarray}
\noindent where $A_s=\prod_{i\in s}\hat{\sigma}_i^x$ and $B_p=\prod_{i\in p}\hat{\sigma}_i^z$ are the star and plaquette operators of the Kitaev model \cite{Kitaev}. The hamiltonian (\ref{H1}) with $\lambda_0=0$ is said to be written in stochastic matrix form \cite{topo1}.\\
\indent The ground state of this hamiltonian can be computed exactly and is given by \cite{topo1}:
\begin{eqnarray}
\ket{gs}&=&\sum_{g \in G}  \frac {e^{-\beta \sum_i \sigma_i^z(g)/2} } {\sqrt{Z(\beta)} } g\ket{0},\\
Z(\beta)&=& \sum_{g \in G}e^{-\beta \sum_i \sigma_{i}^{z}(g) }.
\end{eqnarray}
\noindent Here, $G$ is the abelian group generated by all the star operators $A_s$, $\ket{0}$ is the completely polarized state corresponding to all the spins in the $+1$ eigenstate of  $\sigma^z$ and $\sigma_i^z(g)$ is the $z$-component of the spin at site $i$ in the state $g\ket{0}$.

Let us try to understand the phases of this model. When $\beta=0$, we have the pure Kitaev toric code. Its ground state is a closed string condensed phase. An $x(z)-$string is a collection of spins that are flipped in the $\sigma^z(\sigma^x)$ basis. The term with the plaquette operator says that only closed strings are allowed. The term containing the star operators $A_s$ makes instead closed strings of flipped spins to be created and fluctuate. This phase is topologically ordered, as is shown by a non-vanishing topological entropy \cite{enttopo}-\cite{enttopo3}. We can regard the $\beta$-dependent term as a kind of tension for the $z-$strings. As we increase $\beta$, larger loops are less favored. Indeed one can see that for small $\beta$ the model is the toric code in an external magnetic field. For larger $\beta$ the phase is not topologically ordered, as one can infer from the vanishing of the topological entropy \cite{topo1}. This is why one can use topological entropy as an order parameter \cite{num,topo1,exactw}. One expects that for a particular value of $\beta$ the system undergoes a QPT from the the topologically ordered phase to a ``magnetically" ordered phase. The authors in \cite{topo1} proved that this model has a second order phase transition at $\beta_c=(1/2)\ln(\sqrt{2}+1)$. For $\beta<\beta_c$ the system has a topologically ordered phase, with $S_{\mathrm{topo}}=2$, and for $\beta>\beta_c$ the topological entropy vanishes: $S_{\mathrm{topo}}=0$. It is very important to notice that despite being not topologically ordered, such phase is not a Landau-Ginzburg phase. There is no local order parameter to characterize it \cite{topo1}.

We now analyze this transition using the fidelity between two ground states $\ket{\beta}$ and $\ket{\beta + \delta\beta}$ corresponding to slightly different values of the relevant parameter $\beta$. Therefore, we consider this quantity:

\begin{eqnarray}
F(\beta,\beta+\delta\beta)&=&\langle gs(\beta)| gs(\beta+\delta\beta) \rangle \nonumber \\
&=& \sum_g \frac{e^{-(\beta+1/2\delta\beta)\sum_i\sigma_i^z(g)} }{ \sqrt{Z(\beta)}  \sqrt{Z(\beta+\delta\beta))} }.  \label{fidelity1}
\end{eqnarray}

Expanding (\ref{fidelity1}) to second order in $\delta \beta$ i.e., $F\approx 1- g_{\beta\beta} \delta\beta^2,$ we obtain the following 
{\em fidelity  metric} \cite{F 5} $g_{\beta\beta}$:

\begin{eqnarray}\label{fubini1}
g_{\beta\beta}&=&\frac{1}{4}\Bigg[ \frac{\sum_{g\in G}\big(\sum_i\sigma_i^z(g) \big)^2e^{-\beta\sum_j\sigma_j^z(g)}}{Z(\beta)} \\  \nonumber 
&-& \bigg(\sum_{g\in G}\frac{\sum_i\sigma_i^z(g) e^{-\beta\sum_j\sigma_j^z(g)}}{Z(\beta)} \bigg)^2\Bigg].
\end{eqnarray}
\indent Much in the spirit of the fidelity approach, near the quantum phase transition there is an enhancement in the distinguishability between the ground states $\ket{\beta}$ and $\ket{\beta+\delta\beta}$, resulting in a superextensive scaling of the singular behavior of this metric at the critical point. Indeed, this singular behavior can be captured by mapping this quantum model to a classical statistical model, in the following way. Any group element $g\in G$ is the product of the star operators in some set $\mathcal S (g)$. Thus, $g\ket{0}$ is completely specified by the same set, modulo the product of all such operators, equal to the identity for periodic boundary conditions (i.e., for a torus of genus one). Then, for every two configurations specified by $\{g\in G\}$ there will correspond one configuration $\{\theta\}$ of a classical Ising model with degrees of freedom $\theta_s$ on the sites, such that $\theta_s=-1 (+1)$ when the corresponding star operator $A_s$ is (is not) acting on the site $s$. Since a spin $\sigma_i^z$ can be flipped only by its two neighboring $\theta$ spins, we have that $\sigma_i=\theta_s\theta_s'$, with $s$ and $s'$ the end points of the bond $i$. In that case, defining $E_{\mathrm{Ising}}=J\sum_{<s,s'>}\theta_s\theta_s'$, we obtain $Z_{\mathrm{Ising}}=\sum_{\theta}e^{-\beta\sum_{<s,s'>}\theta_s\theta_s'}=2\sum_{g \in G}e^{-\beta \sum_i \sigma_{i}^{z}(g) } = 2Z(\beta)$, where we took $\beta=J/T$ for the Ising model. Using this equality, we can write (\ref{fubini1}) as:
\begin{eqnarray}
g_{\beta\beta}=\frac{1}{4\beta^2} C_v,
\end{eqnarray} 
where $C_v$ is the specific heat of the 2D Ising model. It is well known that  $C_v$ has a logarithmic divergence at criticality \cite{Baxter}. Hence mapping to the classical Ising model reveals that the fidelity  metric has a logarithmic divergence 
\begin{equation}
g_{\beta\beta}\sim\ln|\beta_c/\beta-1|,
\end{equation}
at $\beta_c=\frac{1}{2}\ln (\sqrt{2}+1)$. \\ 
\indent In \cite{topo1} the authors remark that indeed the phase transition to the topologically ordered phase could be detected by the local magnetization $m(\beta)=\frac{1}{N}\sum_i \langle \hat{\sigma}_i^z\rangle=\frac{1}{N}E_{\mathrm{Ising}}(\beta)$, with its first derivative equal to the specific heat, i.e., $\frac{\partial m}{\partial \beta}=-\frac{1}{N}\beta^2C_{\mathrm{Ising}}(\beta)$, where $N$ is the number of sites. We see that the fidelity metric captures very naturally this divergence since it is equivalent to the specific heat, which diverges at the critical point.

{\em Topological QPT in the quantum eight-vertex model.---} We now turn to analyze another model that exhibits a transition to a topological phase, and in which the mapping to a classical statistical model can be performed to analyze the scaling of the fidelity metric near the critical point. This model is the so-called quantum eight-vertex model, defined and studied in references \cite{8V1} and \cite{8V2}. We proceed to review this model very briefly.\\
\indent The classical eight-vertex model \cite{Baxter} consists of arrows placed along the bonds of a square lattice. The arrows can point in either direction along each of the bonds, subject to the constraint that an even number of arrows go into (and out of) each site. There are eight distinct configurations for the arrows around each site satisfying this constraint. Each vertex configuration is assigned an energy $\epsilon_i$. Furthermore, by imposing toroidal boundary conditions,  symmetry under rotations and inversions of all spins (i.e., zero external electric field), one finds that there are only two independent Boltzmann weights, usually denoted by $c$ and $d$, with $c=e^{-\epsilon_c/T}$ and $d=e^{-\epsilon_d/T}$. The partition function of this model has the form $Z(c,d)=\sum_{\mathcal{C}}c^{n_c(\mathcal{C})}d^{n_d(\mathcal{C})}$, with $n_c(\mathcal{C})$ and $n_d(\mathcal{C})$ the number of $c$ and $d$ type vertices for the configuration $\mathcal{C}$. The total energy for a configuration $\mathcal{C}$ is given by $E=n_c(\mathcal{C})\epsilon_c+n_d(\mathcal{C})\epsilon_d$.

This classical model can be exactly solved in the thermodynamic limit by computing the free energy density using the highest eigenvalue of the transfer matrix \cite{Baxter}. It exhibits ordered phases for $d>c+2$ and $d<c-2$, where the $\mathbb{Z}_2$ symmetry of flipping all the arrows is spontaneously broken, while the system is disordered for $|c-d|<2$. For $d<c-2$, there is a proliferation of $c$ vertices, while for $d>c+2$ the $d$ vertices dominate. These phases are called ``antiferroelectric".\\
\indent One unusual feature of the classical model is that the critical exponents change continuously along the critical lines $d=c+2$ and $d=c-2$, with the free energy density having a singular behavior near $d=c-2$ of the form 
\begin{eqnarray}
f_{\mathrm{sing}}\sim | |d-c|-2 |^{\pi/\mu}, \label{8Vf}
\end{eqnarray}
with $\mu=2\tan^{-1}\sqrt{cd}$. When $\pi/\mu=m$, with $m$ an integer, this expression is changed by an additional logarithmic divergence: $f_{\mathrm{sing}}\sim | |d-c|-2 |^{\pi/\mu}\ln(|d-c|-2)$. The model is also critical along the lines $c=0, d\le2$ and $d=0, c\le2$, since it reduces to the disordered phase of the six-vertex model, which has an infinite correlation length. The points $c=0$, $d =2$ and $d = 2$, $c=0$ are BKT critical points. There, the exponent $\pi / \mu$ diverges.\\ 
\indent The quantum eight-vertex model \cite{8V1} is defined such that its Hilbert space basis $\{\ket{\mathcal{C}}\}$ is given by the configuration space of the classical eight-vertex model, with each state real and orthonormal to each other. The hamiltonian of this model is of the form $H = \sum_{i}Q_i$, with $Q_i$ positive operators, chosen such that $H$ annihilates the following state:
\begin{equation}
\ket{gs(c^2,d^2)}=\frac{1}{\sqrt{Z_{8V}^Q(c^2,d^2)}}\sum_{\{\mathcal{C}\}}c^{\hat{n}_c(\mathcal{C})}d^{\hat{n}_d(\mathcal{C})}\ket{\mathcal{C}},\nonumber
\end{equation}
\noindent with the normalization factor given by: $Z_{8V}^{Q}(c^2,d^2)=\sum_{\{\mathcal{C}\}}c^{2\hat{n}_c(\mathcal{C})}d^{2\hat{n}_d(\mathcal{C})}$, where $\hat{n}_c(\mathcal{C})$ and $\hat{n}_d(\mathcal{C})$ are the number operators for the $c$ and $d$ type vertices, for the configuration $\mathcal{C}$ \cite{8V2}. The authors in \cite{8V1} and \cite{8V2} noted that since the normalization factor above is the partition function for the classical two-dimensional eight-vertex model with weights $c^2$ and $d^2$, then the ground-state phase diagram for the quantum model is identical to the classical one, but given in terms of $c^2$ and $d^2$. The quantum model exhibits a topologically ordered phase in the region of the phase diagram that corresponds to the disordered phase of the classical model. Indeed, the topological entropy in the quantum model is given by $S_{\mathrm{topo}}=-\ln(2)$ in the topological phase $|d^2-c^2|<2$, while it is zero elsewhere. In particular, for $c^2 = d^2 = 1$ one recovers the ground state of the Kitaev model \cite{Kitaev}.\\ 
\indent Let us now pursue a fidelity analysis of this quantum phase transition. Again the mapping to the classical model proves useful. As we will see, the fidelity metric is equal to the fluctuations in the number of $c$ and $d$ type vertices of the classical model. This will provide us with the scaling of the metric near the phase transition. \\
\indent Consider then the fidelity between two ground states for slightly different values of the parameters $c^2$ and $d^2$ and expand it to second order in $c^2$ and $d^2$: $F=\langle gs(c^2,d^2)|gs(c^2 + \delta c^2, d^2 + \delta d^2)\rangle
\approx1 - g_{c^2c^2}(\delta c^2)^2 -g_{c^2d^2}(\delta d^2)^2-g_{c^2d^2}(\delta c^2 \delta d^2)$, where the metric elements of the $2$x$2$ fidelity metric are given by:
\begin{eqnarray}
g_{c^2c^2}&=&\frac{1}{4c^4}\big(\langle n_c^2\rangle - \langle n_c\rangle^2\big),\label{gcc}\\
g_{d^2d^2}&=&\frac{1}{4d^4}\big(\langle n_d^2\rangle - \langle n_d\rangle^2\big), \label{gdd}\\
g_{c^2d^2}&=&\frac{1}{2c^2d^2}\big(\langle n_c n_d\rangle - \langle n_c\rangle\langle n_d\rangle\big), \label{gcd}
\end{eqnarray}
where the averages are now taken with respect to the classical eight-vertex model. 

Using this equivalence with the classical model, we can get the scaling of those metric elements near criticality by using the expression for the free energy density $f=-T\lim_{N \to \infty}N^{-1}\ln Z(c^2,d^2)$ as a generating function for correlations, by differentiating with respect to the energies $\epsilon_c$ and $\epsilon_d$. We obtain then that the dominant scaling near criticality of the metric elements is: 
\begin{equation}
g_{c^2c^2},\ g_{d^2d^2},\ g_{c^2d^2}\sim| |d^2-c^2|-2 |^{\pi/\mu-2},
\end{equation}
and $||d^2-c^2|-2 |^{\pi/\mu-2}\ln{||d^2-c^2|-2 |}$ for $\pi/ \mu$ an integer. Then, we have an algebraic divergence of the fidelity metric only for $\pi/\mu-2<0$, and a logarithmic divergence for $\pi/\mu-2=0$.  Using the fact that near criticality $\mu=2\tan^{-1}\sqrt{c^2d^2}$, those two conditions can be written as $1<c^2d^2$ and $c^2d^2=1$, respectively. Contrary to the case analyzed before,  the metric now diverges as a power law instead of logarithmically, but only for a certain region of the phase diagram.

Some remarks are now due. The eight-vertex model can be shown to be equivalent to two classical square Ising lattices, coupled with a quartic spin term \cite{Baxter}, with both models at the same temperature. It is interesting to note that the curve $c^2d^2=1$ corresponds to the line along which the coupling between the four spins disappears, and separates the region where this coupling is ferromagnetic and anti-ferromagnetic. The region $1<c^2d^2$ corresponds to this last case. \\
\indent Furthermore, given the equivalence of the eight vertex model to a pair of decoupled Ising models for $c^2d^2=1$, one can contemplate the effects of disorder onto the previous scaling results by considering the two dimensional Ising lattice model with random bonds. For $c^2d^2=1$ the metric elements (\ref{gcc})-(\ref{gcd}) are now proportional to the specific heat $C_v$ of the Ising 2D model, which can scale as $\log(\log t)$ at criticality, with $t=(T-T_c)/T_c$ the reduced temperature \cite{Dotsenko}, \cite{Talapov}. We conclude that the fidelity metric elements can present this novel doubly logarithmic divergence for the disordered quantum eight vertex model as well.

\indent Finally, there are many models which are either a special case of the eight-vertex model or equivalent to it, such as the quantum XYZ chain model, ice model, F model, etc \cite{Baxter}. Therefore, by analyzing the fidelity metric in the quantum eight-vertex model we can infer its behavior in many other models and predict its divergence only for regions in the phase diagram that correspond to the condition $1\le c^2d^2$. \\



{\em Conclusions and outlook.---} In this paper we have performed  a fidelity analysis of quantum phase transitions to topologically ordered phases.
We have considered two different systems that can be naturally mapped onto classical statistical mechanical models. The mapping reveals that the fidelity metric corresponds to derivatives of the free energy with respect to some parameters of the model, giving rise to correlations in the classical system. 

We discovered a logarithmic divergence in the fidelity metric for the stochastic matrix form Hamiltonian model near the transition to the topologically ordered state. This may be related to the fact that in this model there is no local oder parameter nor symmetry breaking, and only topological order is involved. In perspective, this is an aspect that deserves more investigation. On the other hand, the quantum eight vertex model still has a power law divergence at the transition to the topological phase, but exhibits this singularity for a restricted region of the phase diagram only.\\
\indent A satisfactory understanding of the relation bewteen fidelity metric singularities and the nature of different topological as well as standard orders
involved in the transitions is a primary goal for future investigations.

{\em Acknowledgements.---}
We would like to thank H. Saleur for suggesting us the disorder case,  L. Campos Venuti for fruitful discussions, and N. Tobias Jacobson for useful comments.  

{\em Note added.---} Upon the completion of this work, we became aware of related results for topological phases in the Kitaev honeycomb model in \cite{F-china} and \cite{G-china}.


\begin{thebibliography}{99}


\bibitem{Wen} X. -G. Wen, \textit{Quantum Field Theory of Many-Body Systems}, Oxford University Press, 2004. 
\bibitem{gold} N. Goldenfeld,  \textit{Lectures on phase transitions and the renormalization group}, Westview Press, Boulder, 1992.
\bibitem{Tsui} D.C.~Tsui, H.L.~Stormer, and A.C.~Gossard, Phys. Rev. Lett. \textbf{48}, 1559 (1982).

\bibitem{Laughlin} R.B.~Laughlin, Phys. Rev. Lett. \textbf{50}, 1395 (1983).
\bibitem{enttopo} A. Hamma, R. Ionicioiu and P. Zanardi, Phys. Rev. A {\bf{71}}, 022315 (2005).
\bibitem{enttopo2} A. Kitaev and J. Preskill, Phys. Rev. Lett {\bf{96}}, 110404 (2006).
\bibitem{enttopo3} M. Levin and X. -G. Wen, Phys. Rev. Lett {\bf{96}}, 110405 (2006). 
\bibitem{QPT} S. Sachdev, \textit{Quantum Phase Transitions}, Cambridge University Press, Cambridge, England, 1999. 

\bibitem{F 0} H. T. Quan, Z. Song, X. F. Liu, P. Zanardi and C. P. Sun, Phys. Rev. Lett. {\bf 96}, 140604 (2006).
\bibitem{F 1} P.~Zanardi and N.~Paunkovi\' c, Phys. Rev. E {\bf 74}, 031123 (2006).
\bibitem{F 3} W. -L. You, Y. -W. Li and S. -J. Gu, Phys. Rev. E {\bf{76}}, 022101 (2007).
\bibitem{F 4} S. Chen, L. Wang, S. -J. Gu and Y. Wang, Phys. Rev. E {\bf{76}}, 061108 (2007).
\bibitem{F 5}P. Zanardi, P. Giorda and M. Cozzini, Phys. Rev. Lett. {\bf{99}}, 100603 (2007).
\bibitem{F 6} L.  Campos Venuti and P. Zanardi, Phys. Rev. Lett. {\bf{99}}, 095701 (2007).
\bibitem{F 7} P. Zanardi, M. Cozzini and P. Giorda, J.~Stat.~Mech.~L02002 (2007).
\bibitem{F 8} M. Cozzini, P. Giorda and P. Zanardi, Phys.~Rev.~B \textbf{75}, 014439 (2007).
\bibitem{F 9}  M.~Cozzini, R.~Ionicioiu and P.~Zanardi, Phys. Rev. B {\bf{76}}, 104420 (2007).
\bibitem{F 10} P.~Zanardi, H.-T.~Quan and X.-G.~Wang and C.-P.~Sun, Phys. Rev. A {\bf{75}}, 032109 (2007).
\bibitem{F 11} P. Buonsante and A. Vezzani, Phys.~Rev.~Lett.~\textbf{98}, 110601 (2007).
\bibitem{F 12} P. Zanardi, L. Campos Venuti and P. Giorda,  in Phys. Rev. A {\bf{76}}, 062318 (2007).
\bibitem{za-paris} P. Zanardi and M.G.A Paris, arXiv:0708.1089.
\bibitem{num} A. Hamma, W. Zhang, S. Haas and D.A. Lidar, arXiv:0705.0026 .
\bibitem{F 13} H. -Q. Zhou, J. -H. Zhao and B.Li, arXiv:0704.2940.
\bibitem{F 14} H.-Q. Zhou, arXiv:0704.2945.
\bibitem{F 15} H. -Q. Zhou, J. -H. Zhao, H. -L. Wang and B. Li, arXiv:0711.4651.
\bibitem{F 16} M. -F. Yang, Phys. Rev. B {\bf{76}}, 180403(R) (2007).
\bibitem{F 17} Y. -C. Tzeng and M. -F. Yang, Phys. Rev. A {\bf{77}}, 012311 (2008).
\bibitem{F 18} H. -Q. Zhou, arXiv: 0803.0585v1.
\bibitem{F 19} H.-Q. Zhou and J. P. Barjaktarevi\v c, arXiv:cond-mat/0701608.
\bibitem{F 20} S. Chen, L. Wang, Y. Hao and Y. Wang, arXiv: 0801.0020.
\bibitem{F 21} S. -J. Gu, H. -M. Kwok, W. -Q. Ning and H. -Q. Lin, arXiv:0706.2495.
\bibitem{F 22} N. Paunkovi\'c, P. D. Sacramento, P. Nogueira, V. R. Vieira and V. K. Dugaev, arXiv: 0708.3494.
\bibitem{F 23} N. Paunkovi\'c and V. R. Vieira, Phys. Rev. E {\bf{77}}, 011129 (2008).

\bibitem{SMD} C. Castelnovo, C. Chamon, C. Mudry and P. Pujol, Ann. Phys. (NY) {\bf{318}}, 316 (2005).
\bibitem{8V1} E. Ardonne, P. Fendley and E. Fradkin, Ann. Phys. {\bf{303}}, 493 (2004).
\bibitem{TNrep} H.-Q. Zhou, R. Orus and G. Vidal, Phys. Rev. Lett. {\bf 100}, 080601 (2008).
\bibitem{topo1} C. Castelnovo and C. Chamon, Phys. Rev. B {\bf 77}, 054433 (2008). 
\bibitem{Kitaev} A. Y. Kitaev, Ann. Phys. (N.Y.) {\bf{303}}, 2 (2003).
\bibitem{exactw} J.~Yu, S.-P.~Kou and X.-G.~Wen, arXiv:0709.2276.
\bibitem{8V2} S. Papanikolaou, K. S. Raman and E. Fradkin, Phys. Rev. B, {\bf{76}}, 224421 (2007).
\bibitem{Baxter} R. J. Baxter,  \textit{Exactly Solved Models in Statistical Mechanics}, Academic Press, 1982.
\bibitem{Dotsenko} Vik S Dotsenko and Vl S Dotsenko, J. Phys. C., {\bf{15}}, 495 (1982).
\bibitem{Talapov} A. L. Talapov and L. N. Shchur, J. Phys.: Condens. Matter {\bf{6}}, 8295 (1994).
\bibitem{F-china} J.-H. Zhao and H.-Q. Zhou, arXiv:cond-mat/0803.0814v1.
\bibitem{G-china} S. Yang, S. -J. Gu, C. -P. Sun and H. -Q. Lin, arXiv:0803.1292v1.
\end{thebibliography}
\end{document}